\documentclass[a4paper,english,aps,pra,showpacs,floatfix,nofootinbib,twocolumn]{revtex4}
\usepackage{mathptmx}
\usepackage[T1]{fontenc}
\usepackage[latin9]{inputenc}
\usepackage{color}
\usepackage{babel}
\usepackage{amstext}
\usepackage{amssymb}
\usepackage{graphicx}
\usepackage{esint}
\usepackage[unicode=true,pdfusetitle,
 bookmarks=true,bookmarksnumbered=false,bookmarksopen=false,
 breaklinks=true,pdfborder={0 0 0},pdfborderstyle={},backref=false,colorlinks=true]
 {hyperref}
\usepackage{breakurl}

\makeatletter

\@ifundefined{textcolor}{}
{%
 \definecolor{BLACK}{gray}{0}
 \definecolor{WHITE}{gray}{1}
 \definecolor{RED}{rgb}{1,0,0}
 \definecolor{GREEN}{rgb}{0,1,0}
 \definecolor{BLUE}{rgb}{0,0,1}
 \definecolor{CYAN}{cmyk}{1,0,0,0}
 \definecolor{MAGENTA}{cmyk}{0,1,0,0}
 \definecolor{YELLOW}{cmyk}{0,0,1,0}
}

\@ifundefined{showcaptionsetup}{}{%
 \PassOptionsToPackage{caption=false}{subfig}}
\usepackage{subfig}
\makeatother

\begin{document}

\title{Light-matter interaction of a quantum emitter near a half-space graphene nanostructure}

\author{Vasilios Karanikolas$^{a}$, Pelin Tozman$^{b}$ and Emmanuel Paspalakis$^{a}$}

\affiliation{$^{a}$Materials Science Department, School of Natural Sciences,
University of Patras, Patras 265 04, Greece, $^{b}$Physics Engineering
Department, Hacettepe University, Ankara, Turkey}

\date{\today }
\begin{abstract}
The Purcell factor and the spontaneous emission spectrum of a quantum
emitter (QE) placed close to the edge of a graphene half-space nanostructure
is investigated, using semi-analytical methods at the electrostatic
regime. The half-space geometry supports an edge and a bulk surface
plasmon (SP) mode. The Purcell factor of the QE is enhanced over
eight orders of magnitude when its emission energy matches the resonance
energy modes, for a specific value of the in-plane wave vector, at
a separation distance of $5\,$nm. The different transition dipole
moment orientations influence differently the enhancement factor of
a QE, leading to large anisotropic behavior when positioned at different
places above the half-space geometry. The field distribution is presented,
showing clearly the excitation of the SP modes at the edge of the
nanostructures. Also, we present the spontaneous emission spectrum
of the QE near the half-space graphene nanostructure and show that strong light-matter coupling may emerge.
When a QE with a free-space lifetime of $1\,$ns is placed at a distance of $10\,$nm away from the edge of the graphene half-space, a Rabi splitting
of $80 \,m$eV is found. Our contribution can be used for designing
future quantum applications using combination of QEs and graphene nanostructures.
\end{abstract}
\maketitle

\section{INTRODUCTION\label{sec:I}}

\begin{figure}[t]
\includegraphics[width=0.4\textwidth]{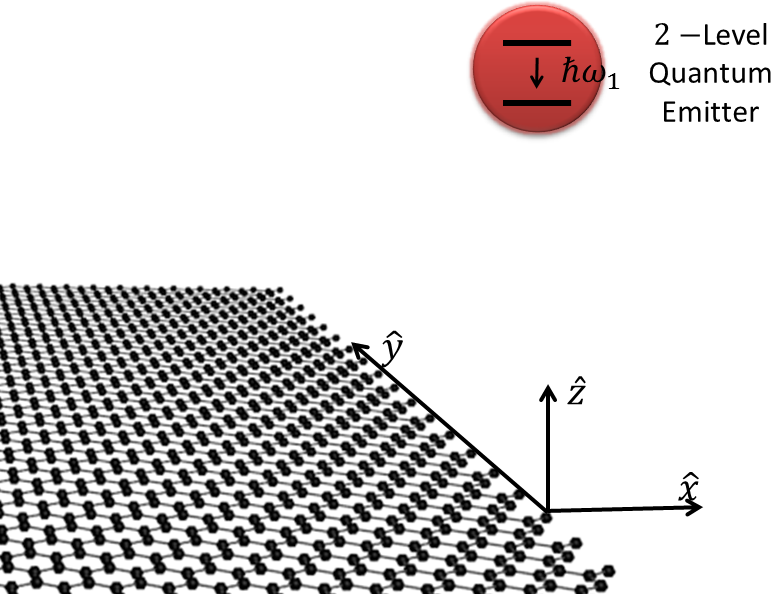}\caption{A QE placed above a graphene half-space geometry, the nanostructure
has been placed in the $x-y$ plane, $x<0$. \label{fig:Fig01}}
\end{figure}

The polariton modes, which are collective hybrid light-matter modes,
open the opportunity to control light at the nanoscale \cite{Basov16a,Low2017}.
Among the various hybrid modes the most studied ones are the surface
plasmon (SP) modes. The SP are hybrid modes of the electromagnetic
field and the conduction band electrons, which are confined in the
interface between a metal and a dielectric, where they propagate along
it and are confined in the perpendicular dimension. Graphene is a
material with higher optical and mechanical properties that is also
expected to replace noble metals (Au and Ag mainly) in photonics applications
based on SP modes, due to lower material losses, at the mid-
to far- infrared parts of the spectrum \cite{Low2014a,GarciadeAbajo2014}.
The graphene SP excitation modes are in the infrared part of the
spectrum \cite{Koppens2011}, and are tunable, controlled by electric
field gating, doping or multilayer stacking. Also, the fabrication capabilities
and detection technologies developed over the last decade open new
routes for manipulating the light to the extreme for building sophisticated
applications \cite{Fang2013,Tielrooij2015}.

The SP modes can be excited by different ways, such as plane wave,
fast electrons or localized sources, like quantum emitters (QEs), which
have active properties \cite{GarciadeAbajo2010a}. For infinite planar
graphene nanostructures special techniques are needed to excite the
SP modes due to the large miss-match in the momentum between them and the incoming
light \cite{Nikitin2014}. On the same time localized scatterers have been
used to launch surface plasmon modes to localized and extended graphene
nanostructures, like Au particles \cite{Alonso2014} or metallic tips
\cite{Chen2012,Fei2011}, where the scattered field can provide the
high wave vector components needed to excite the SP modes. Novel experimental
techniques have also been used to launch the SP modes and also
to measure the fields amplitude and phase \cite{Gerber2014}. However,
these techniques are limited by the excitation wavelength and
do not operate at the near infrared part of the spectrum.

Surface polariton modes, of any kind plasmon, exciton or phonon, can
confine light at the extreme \cite{Basov16a,Low2017} thus opening
new routes for detection application, but also for applications for
novel quantum technologies. For these all-optical devices various
processes need to be handled on-chip, where the basic idea is to be
able to transfer to the chip information and extract from the
chip its response. The information can be transmitted/extracted to/from
this chip by a QE. A QE is described as a two-level system, approaching
various physical systems, such as atoms, molecules, quantum dots,
quantum wells and color centers \cite{Baranov2018}.
So, we need to fully understand
how a QE interacts with a device at its boundary.

We note that the interaction of
QEs with graphene monolayers and graphene nanostructures, like graphene waveguides and nanodisks, is a topic of active research \cite{Koppens2011,Cox12a,Manjavacas12a,Forati14a,Sun2014,Karanikolas2015,Karanikolas16b,
Cuevas2016,Rivera16a,Chang17a,Sun2017,Cuevas18,Sloan18a,Karanikolas18a,
Zhang18a,Sanders18a,Fang18a,Cox18a,Cuevas19a,Zeng19a,Zhang19a}.
However, the interaction of a QE with a graphene half-space geometry, to the best of our knowledge, remains unexplored. Thus, here we investigate
the spontaneous emission (SE) properties of a QE placed near the edge of
a graphene half-space geometry, which is depicted at Fig.~\ref{fig:Fig01}.
Furthermore, we study the edge and bulk SP modes and present the SP
field intensity launched by a QE.
The QE/graphene half space geometry interaction is described using
the non-Hermitian quantum description of light-matter interaction
\cite{Dung2000}. In the heart of this theory is the knowledge of
the electromagnetic (EM) Green's tensor \cite{Nanoopticsbook}, which has a classical interpretation,
it gives the electromagnetic response of the geometry under consideration
to a point-like dipole excitation. Here, we calculate this quantity
in the electrostatic regime, since the graphene half-space geometry
can support SP modes up to the near infrared part of the spectrum
\cite{Fetter1986,Wang2011}, and extract from this the Purcell factor. 

The Purcell factor of the QE is the main quantity of
interest in the weak light-matter coupling regime, as it quantifies the change of the SE rate of the QE. We find that the Purcell factor takes very large values, larger than $10^8$, near the edge of the graphene half-space geometry.
We also show that when the QE/graphene half-space nanostructure
separation is small, we enter the strong light-matter coupling regime \cite{Baranov2018,Vanvlack12a,Tejedor14,Hakami14,Li16a,Thanopulos2017,Rousseaux18a,Karanikolas19a}.
At this regime the combined system QE/half-space geometry coherently
exchange energy between its two constituent elements. The strong coupling
regime appears as a splitting, the Rabi splitting, in the SE spectrum of the
QE near the graphene half-space nanostructure.

We start in Sec.~\ref{subsec:IIa} by presenting the mathematical
frame under which we calculate the Purcell factor and the SE spectrum of a QE, using the
non-Hermitian description of the light-matter interaction. In Sec.~\ref{subsec:IIb-1}, where
the dispersion relation, the propagation length and the penetration
depth of the graphene half-space are analyzed, we identify the edge
and bulk SP modes. In Sec.~\ref{sec:III} we present the Purcell
factor of a QE located at the edge of the graphene half-space nanostructure placed at the $xy$ plane,
while varying its emission energy and the in-plane wave vector value,
for the $z-$ and $x-$ transition dipole moments of the QE. Also,
for specific values of the emission energy and of the wave vector
we vary the the position of the QE along the nanostructure, in order
to probe the behavior of the dipole orientation. The Green's tensor
amplitude, which is connected with the EM field, created by a QE is
also presented for the edge and bulk SP modes. This section is finalized by presenting the SE spectrum of a QE, placed at the edge of the graphene half-space geometry. A pronounced Rabi splitting is achieved, showing that the QE/half-space interaction operates at the strong coupling regime. Sec.~\ref{sec:IV}
concludes our findings. In Appendix ~\ref{sec:Appendix-A}
we present an expansion method, to a known set of functions, to calculate
the Green's tensor for the case of the graphene half-space geometry
at the electrostatic limit. In Appendix ~\ref{sec:Appendix-B} we present
the surface conductivity of graphene, that gives its optical response.

\section{Theoretical description of the light-matter interactions\label{sec:II}}

\subsection{Purcell factor and spontaneous emission rate\label{subsec:IIa}}

As localized sources we consider QEs which are
approximated as two-level systems. The ground state of the QE is $|g\rangle$ and the excited state
is $|e\rangle$. The transition frequencies from the excited to the ground
state and the dipole matrix element are denoted as $\omega_{1}$
and $\mathbf{\mu}$, respectively. An initially excited QE
interacts with its environment through the electromagnetic field and
relaxes from its excited state to the ground state by emitting a photon
(radiative relaxation path) or exciting any of the dressed states
supported by its environment. For the case investigated here the
dressed states will be the edge and bulk SP modes. The initial state
of the system is denoted as $|i\rangle=|e\rangle\otimes|0\rangle$,
where the QE is in the excited state and the electromagnetic field
is in its vacuum state. The QE will relax to the final state of the
entire system $|f\rangle=|g\rangle\otimes\hat{f}_{i}^{\dagger}(\mathbf{r},\omega)|0\rangle$.
By applying Fermi's golden rule and summing over all final states,
the expression for the relaxation rate is $\Gamma(\mathbf{r},\omega_{1})=2\omega^{2}\mu^{2}/\left(\hbar\varepsilon_{0}c^{2}\right)\hat{\mathbf{n}}\cdot\textrm{Im}\mathfrak{G}(\mathbf{r},\mathbf{r},\omega)\cdot\hat{\mathbf{n}}$,
where $\mathbf{\hat{n}}$ is a unit vector along the direction of
the transition dipole moment, $\mu$, and $\mathfrak{G}(\mathbf{r},\mathbf{s},\omega)$
is the EM Green's tensor.

For the case of graphene half-space the Green's tensor for a QE that
his transition dipole moment is along $z$ has the form:
\begin{equation}
\mathfrak{G}_{zz}^{\text{Ind}}(z_{1},\omega)=\frac{1}{k_{0}^{2}}\sum_{j=1}^{\infty}(-1)^{j}c_{j}(z_{1})\int_{-\pi/2}^{\pi/2}d\theta\,\frac{1}{\cos\left(\theta\right)}F(z_{1},\theta),\label{eq:01}
\end{equation}
where the QE is placed at $\mathbf{r}=(0,0,z_{1})$, above the edge
of the graphene half-space, and $c_{j}$ are expansion coefficients
that depend on the position of the QE. For the case we have a QE with a $x$-oriented
transition dipole moment, the EM Green's tensor has the form:
\begin{equation}
\mathfrak{G}_{xx}^{\text{Ind}}(z_{1},\omega)=\frac{-i}{k_{0}^{2}}\sum_{j=1}^{\infty}(-1)^{j}c_{j}(z_{1})\int_{-\pi/2}^{\pi/2}d\theta\,\tan\left(\theta\right)F(z_{1},\theta).\label{eq:02}
\end{equation}
In the above expressions we define $F(z_{1},\theta)=e^{-z_{1}/\cos\theta}e^{i(2j+1)\theta}$,
more details for the calculation are given in Appendix ~\ref{sec:Appendix-A}.

Using Eqs.~(\ref{eq:01}) and (\ref{eq:02}) we obtain the Purcell factor of the QE
\begin{equation}
\tilde{\Gamma}_{i}(\omega_{1},\mathbf{r})=\frac{\Gamma_{i}(\omega_{1},\mathbf{r})}{\Gamma_{0}(\omega_{1})}=\sqrt{\varepsilon}+\frac{6\pi c}{\omega}\hat{n}_{i}\mathrm{Im}\,\mathfrak{G}_{ii}^{\text{Ind}}(\mathbf{r},\mathbf{r},\omega)\hat{n}_{i}\label{eq:03}
\end{equation}
which shows in the weak coupling regime if the SE rate of the QE is enhanced or inhibited in the presence of the graphene
half-space geometry in respect to the free space case. Here, $\varepsilon$ is the
permittivity of the host medium and $\Gamma_{0}$ is the Einstein
$A$-coefficient, with $\Gamma_{0}(\omega_{1})=\omega^{3}_{1}\mu^{2}/3\pi c^{3}\hbar\varepsilon_{0}$.
We stress that we deal with the near field regime of the QE, which means that the
QE - graphene half-space geometry separation is much smaller than
the emission wavelength of the QE, $\left|\mathbf{r}\right|\ll\lambda$.

For the strong coupling a different approach is needed. We need to
describe the full problem where the strength of energy exchange between
the QE/nanostructure combined system is determined through the spectral
density $J(\omega,\mathbf{r})$, which also depends on the Purcell factor, and has the expression \cite{Vanvlack12a,Tejedor14,Hakami14,Li16a,Thanopulos2017,Rousseaux18a,Karanikolas19a}
\begin{equation}
J_{i}(\omega_{1},\omega,\mathbf{r})=\frac{\Gamma_{0}(\omega_{1})}{2\pi}\tilde{\Gamma}_{i}(\omega,\mathbf{r})\left(\frac{\omega}{\omega_{1}}\right)^{3},\label{eq:04}
\end{equation}
with $i=z,x$. We observe that the higher the Purcell factor the stronger the coupling of
the QE with its environment. Also, the value of the dipole strength
is very important for approaching the strong coupling limit, the higher
its value the less enhancement of the Purcell factor is needed \cite{Baranov2018}.
The light emitted spectrum of the QE is given by $S(\omega,\mathbf{r})=\frac{1}{2\pi}\int_{0}^{\infty}dt_{2}\int_{0}^{\infty}dt_{1}e^{-i\omega(t_{2}-t_{1})}\left\langle \hat{\mathbf{E}}^{(-)}(t_{2},\mathbf{r})\cdot\hat{\mathbf{E}}^{(+)}(t_{1},\mathbf{r})\right\rangle $
where, after carrying out the calculations, the full expression has
the form \cite{Vanvlack12a,Hakami14,Karanikolas19a}
\begin{equation}
S(\omega,\mathbf{r})=\frac{1}{2\pi}\left|\frac{\frac{\mu^{2}\omega^{2}}{\varepsilon_{0}c^{2}}\mathbf{\hat{n}}\cdot\mathfrak{G}(\omega,\mathbf{r},\mathbf{r}_{d})}{\omega_{1}-\omega-\int_{0}^{\infty}d\omega^{\prime}J(\omega_{1},\omega^{\prime},\mathbf{r})\frac{1}{\omega^{\prime}-\omega}}\right|^{2} \, . \label{eq:05}
\end{equation}
Here, $\mathbf{r}_{d}$ is the position of the signal detection, $\mathbf{r}$
is the position of the QE and $\omega$ is the emission frequency.

\subsection{Analysis of the surface plasmon modes\label{subsec:IIb-1}}

\begin{figure*}
\subfloat[\label{fig:Fig02a}]{\includegraphics[width=0.3\textwidth]{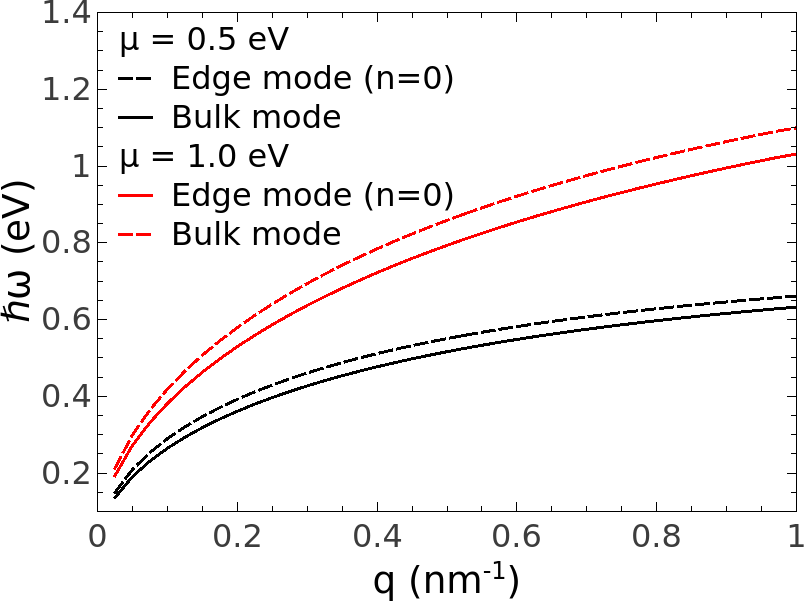}

}~~~\subfloat[\label{fig:Fig02b}]{\includegraphics[width=0.3\textwidth]{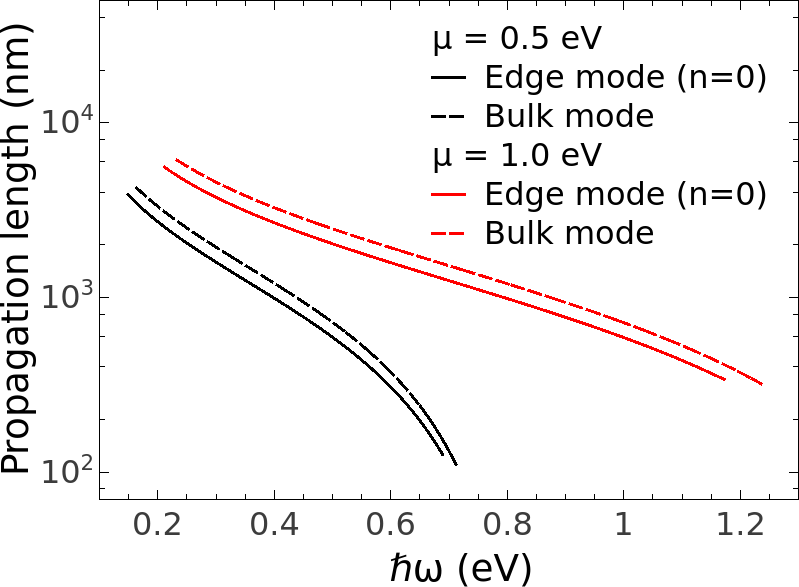}

}~~~\subfloat[\label{fig:Fig02c}]{\includegraphics[width=0.3\textwidth]{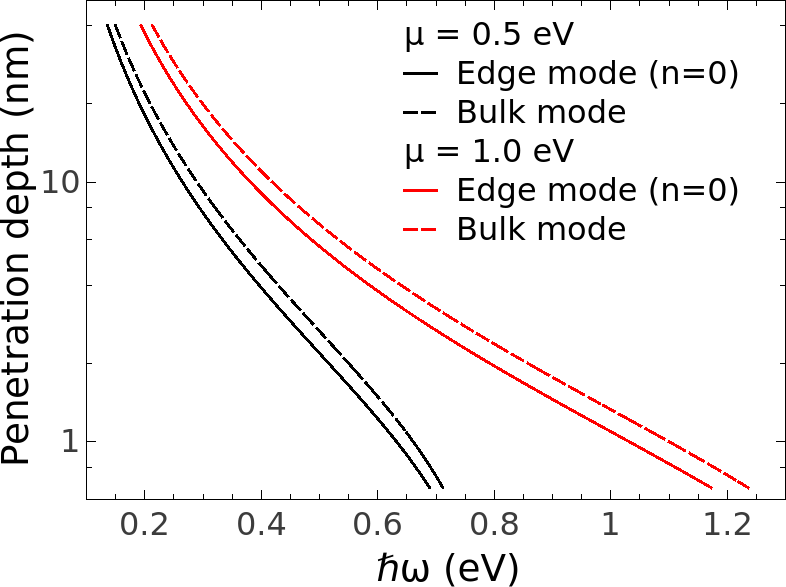}

}

\caption{(a) Dispersion relation, $\hbar\omega(q)$, for a graphene half space
geometry, $\mu=0.5\,$eV and $1.0\,$eV, presenting the $n=0$ and
$1$ edge and bulk modes, respectively. (b,c) The propagation lengths
and the penetration depths of the bulk and edge SP modes, for the
material parameters presented in (a).\label{fig:Fig02}}
\end{figure*}

In Fig.~\ref{fig:Fig02a} we present the dispersion relation $\hbar\omega_{n}(q)$
of the half-space geometry, where we observe that two modes are supported,
the edge and the bulk. To find these resonance frequencies, for each
wave number, we set to zero the external excitation to Eq.~(\ref{eq:A10})
and we numerically solve the equation
\begin{equation}
\frac{\omega_{n}}{\sigma\left(\omega_{n}\right)}=\frac{4iq\zeta_{n}}{\varepsilon},\label{eq:06}
\end{equation}
where $\zeta_{n}$ are the geometric eigenmodes which are solutions
of the equation $\mathbf{K}\mathbf{c}=\zeta\mathbf{Gc}$ and once
and for all are determined for the half space geometry. The exact
forms of the $\mathbf{K}$ and $\mathbf{G}$ matrices are given in the
Appendix~\ref{sec:Appendix-A}, and are obtained via an expansion
over the Laguerre polynomials \cite{Fetter1986}. The optical
response of graphene is described through its surface conductivity $\sigma(\omega)$,
where more details are given in Appendix~~\ref{sec:Appendix-B}. The $n=0$
is connected to the edge mode and the $n=1$ corresponds to the bulk
mode. The modes for $n>1$ they have the same dispersion relation
as the $n=1$ mode and are physically connected with the fact
that the bulk mode is continuous, modified from the bulk mode of the
infinite planar graphene sheet \cite{Karanikolas2015}. In Fig.~\ref{fig:Fig02a}
we observe that the bulk SP mode energy, $\hbar\omega$, is blue shifted
for any given value of the wave vector $q$ compared to the edge SP
mode. All branches are well below the free space light-line, which
is very close to the $y$ axis of Fig.~\ref{fig:Fig02a}.
Furthermore, we observe that by increasing the value of the chemical
potential $\mu$ the dispersion relations for the edge and bulk modes
is blue-shifted a clear indication of the capabilities offered by
graphene for designing tunable photonic applications.

The SP mode is characterized by the propagation length, $L_{\text{SP}}$,
and the penetration depth, $\delta_{\text{SP}}$, supported by the
graphene half-space geometry. The propagation length is connected
with the distance along the graphene half-space geometry that the SP
mode intensity propagates until attenuated to $1/e$ value and it
can be defined as $L_{\text{SP}}=\tau v_{g}$ \cite{Christensen2012},
where $\tau=1\,$ps is the relaxation time of the surface conductivity,
$\sigma_{\text{intra}}$ in Eq.~(\ref{eq:A24a}), that in turn is connected
with the SP modes to acquire a finite lifetime. The relaxation time,
$\tau$, is influenced by several factors, such as collisions with
impurities, coupling to optical phonons, finite-size
and edge effects. Here, $v_{g}$ is the group velocity of the SP
modes and is given by the slope of the dispersion relations $v_{g}=d\omega/dq$
for the edge and bulk modes. The penetration depth is connected with
the extend of the SP mode in the $z$ direction. It physically shows
with how effectively can a QE excite the SP mode by coupling with
his near field regarding the QE's position. Large penetration depths
can be used to effectively couple a QE at larger distances. The penetration
depth is defined as $\delta_{\text{SP}}=1/\text{Im}\left(k_{z}^{\text{SP}}\right)$,
where $k_{z}^{\text{SP}}=\sqrt{k_{0}^{2}-q^{2}}\eqsim iq$ \cite{Karanikolas2015}.

In Fig.~\ref{fig:Fig02b} we present the propagation length, $L_{\text{SP}}$,
for varying energy $\hbar\omega$, considering two values for the
chemical potential $\mu=0.5\,$eV and $1.0\,$eV. We have two propagation
lengths $L_{\text{SP}}^{0}$ and $L_{\text{SP}}^{1}$ for the edge
and the bulk SP modes, respectively. We observe that for any given energy, $\hbar\omega$,
the propagation length for the bulk mode is larger than the propagation
length for the edge mode, for both chemical potential values. Furthermore,
as the energy of the mode is increased, the propagation lengths $L_{\text{SP}}^{0}$
and $L_{\text{SP}}^{1}$ decrease rapidly, because the SPs have sufficient
energy to generate electron-hole pairs and the dispersion relation
is dominated by the interband contributions on the surface conductivity
$\sigma(\omega)$. Moreover, as the chemical potential value $\mu$
is increased the propagation lengths of edge, $L_{\text{SP}}^{0}$,
and bulk modes, $L_{\text{SP}}^{1}$, are also increased. In this case, the SP modes
acquire more energy for propagating along the half-space geometry.
In Fig.~\ref{fig:Fig02c} we observe a similar behavior for the penetration
depth, for the edge, $\delta_{\text{SP}}^{0}$, and bulk, $\delta_{\text{SP}}^{1}$,
modes, with the propagation length, regarding the tunability and the
reduction of their value for increasing energy. Also, the edge mode
is more tightly confined to the graphene half-space geometry than
the bulk mode, since $\delta_{\text{SP}}^{0}<\delta_{\text{SP}}^{1}$
for any given energy for the same chemical potential values. The main
difference between the two lengths is that the propagation lengths
are around two order of magnitudes larger that the penetration depths,
meaning the the SP modes are tightly confined to the graphene half-space
geometry and are able to propagate several hundreds of nanometers
until being attenuated. This properties are important for designing
future practical applications using graphene as a building block.

\section{Emission properties at the weak and strong coupling regimes\label{sec:III}}
In this section we focus on the SE properties of a QE in the presence
of the graphene half-space geometry. High values of the Purcell factor
are presented. The QE can efficiently excite edge and bulk SP modes
by relaxing at the edge of the graphene half-space geometry. Also, strong light-matter coupling is presented in certain cases.
\begin{figure*}
\subfloat[\label{fig:Fig03a}]{\includegraphics[width=0.45\textwidth]{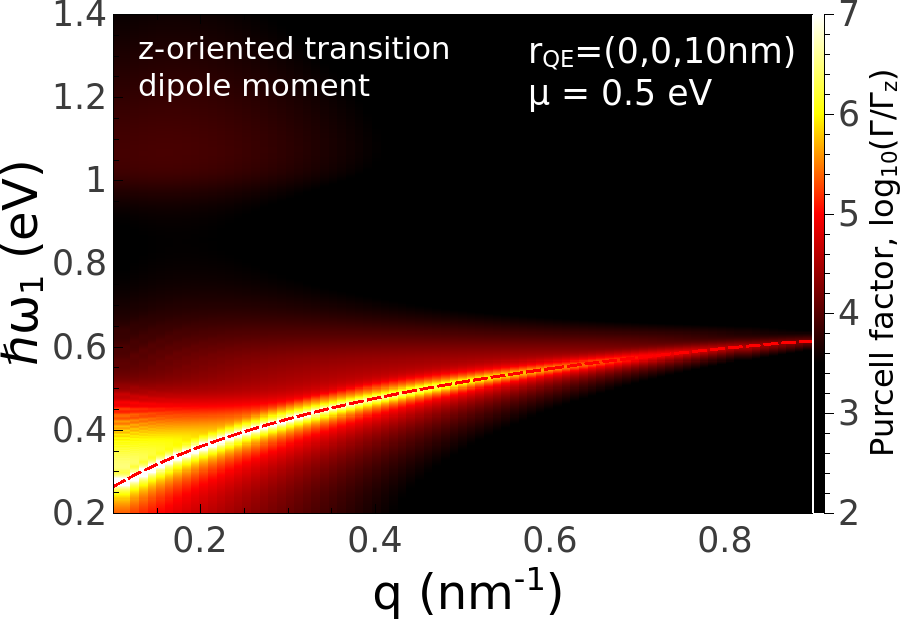}

}\subfloat[\label{fig:Fig03b}]{\includegraphics[width=0.45\textwidth]{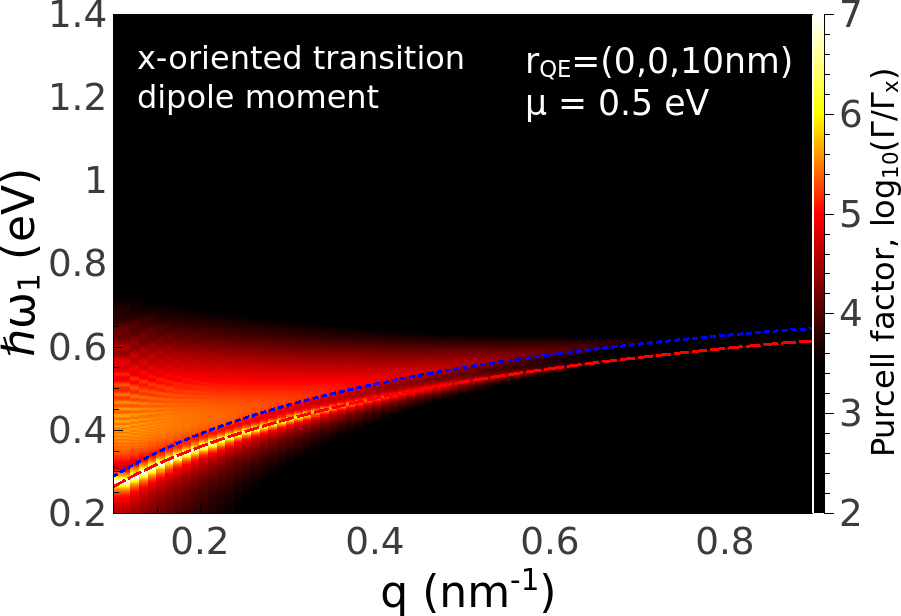}

}

\caption{Contour plot of the Purcell factor of a QE, placed at $\mathbf{r}=(0,0,10\,\text{nm})$,
for varying its emission energy, $\hbar\omega$, for different values
of the in-plane wave vector, $q$. The graphene half-space has a chemical
potential value of $\mu=0.5\,$eV. Two different transition dipole
moments are presented (a) along $z$ and (b) along $x$. The dispersion
relation from Fig.~\ref{fig:Fig02a} are superimposed to the contour
plots.\label{fig:Fig03}}
\end{figure*}

In Fig.~\ref{fig:Fig03} we present a contour plot of the logarithm
of the Purcell factor of a QE, placed at $\mathbf{r}=(0,0,10\,\text{nm})$,
varying  $\hbar\omega_{1}$, for different values of
the in-plane wave vector, $q$. The chemical potential of the graphene
half-space is $\mu=0.5\,$eV and the position of the QE is exactly
at the edge of the graphene half-space nanostructure, see Fig.~\ref{fig:Fig01}.
Two different transition dipole moment orientations of the QE are
considered, in (a) $z$-oriented and in (b) $x$-oriented. We observe
that for a QE with $z$-oriented dipole moment can couple efficiently with
the edge resonance mode, $n=0$. Thus, for these values the Purcell
factor increases above $8$ orders of magnitude in the presence of the graphene half-space structure.
Also, for the specific QE position,
the near field of the QE can efficiently couple with the bulk SP mode,
$n=1$, when its transition dipole moment is oriented along $x$.
Hence, the enhancement of the Purcell factor is above $7$ orders
of magnitude compared to the free space value.
For both orientations, we observe that for increasing the value of
the in-plane wave vector $q$ the Purcell factor of the QE drops.
This has a physical explanation connected with the SP penetration
depth $\delta_{\text{SP}}$ where for the higher $q$ mode the SP
field is more confined to the graphene half-space making it more difficult
for the QE to efficiently excite the SP modes. Thus, for exciting
the higher $q$ modes the QE needs to be placed very close to the
nanostructure to provide the necessary high in-plane wave vector $q$
component values.

\begin{figure*}
\subfloat[\label{fig:Fig04a}]{\includegraphics[width=0.45\textwidth]{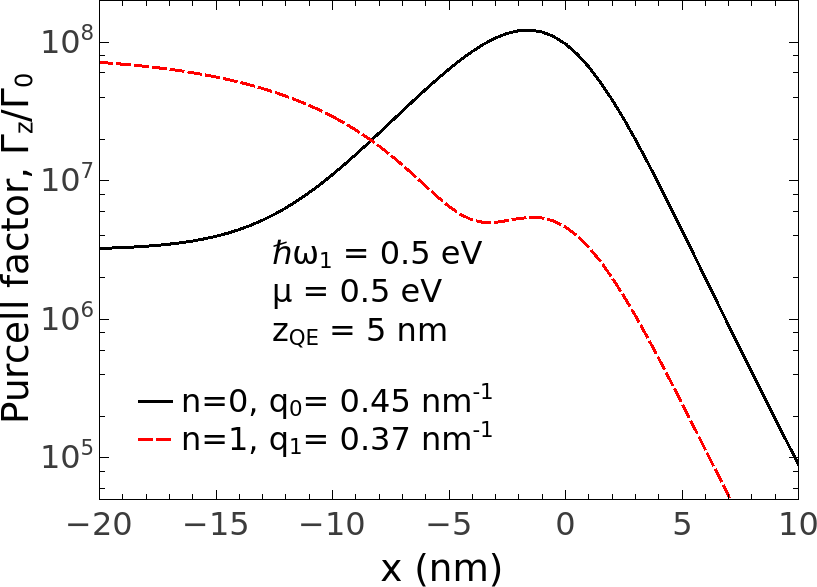}

}\subfloat[\label{fig:Fig04b}]{\includegraphics[width=0.45\textwidth]{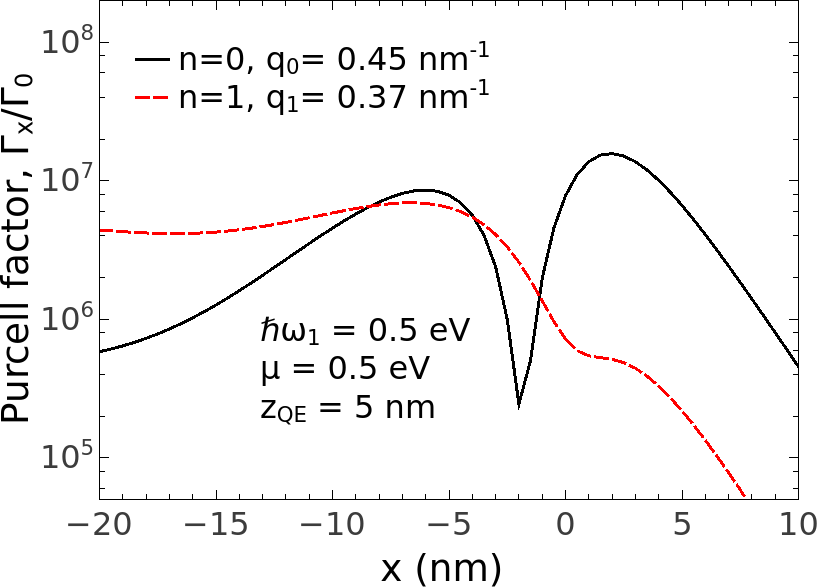}

}

\caption{Purcell factor of a QE above a graphene half-space, with a chemical
potential value of $\mu=0.5\,$eV, varying is position $\mathbf{r}=(x,0,5\,\text{nm})$.
The emission energy of the quantum emitter is fixed at $\hbar\omega_{1}=0.5\,$eV.
The edge, $n=0$, and the bulk, $n=1$, SP contributions are considered,
more details can be found at the insets. Two different transition
dipole moment orientations are considered (a) along $z$ and (b) along
$x$. \label{fig:Fig04}}

\end{figure*}

In Fig.~\ref{fig:Fig04} we present the Purcell factor of a QE when
we vary its position above the graphene half-space geometry, with
chemical potential of $\mu=0.5\,$eV. The QE position is at $\mathbf{r}=(x,0,5\,\text{nm})$
and the emission energy value $\hbar\omega_{1}=0.5\,$eV, matching the
resonance wave vectors $q_{n}$, $n=0$ and $1$ at $0.45\,\text{nm}^{-1}$
and $0.37\,\text{nm}^{-1}$, respectively. The wave vector $q_{0}$
is connected with the edge SP mode and the $q_{1}$ is connected with
the bulk mode. Two transition dipole moment orientations of the QE
are considered, along $z$ in Fig.~\ref{fig:Fig04a} and along $x$
in Fig.~\ref{fig:Fig04b}. We observe in Fig.~\ref{fig:Fig04a}
that the enhancement of the Purcell factor has the largest value for the $n=0$
edge SP mode compared to the bulk mode, more than one order of magnitude
difference, when the QE is close to the edge of the graphene half-space
geometry. As the position of the QE is moved away from the edge of the
graphene half-space to the bulk, $x<0$, the $n=0$ edge mode contribution
to the Purcell factor of the QE is reduced and the bulk SP mode is
enhanced, that is the main relaxation path of the QE. Also, when the
QE is placed out of the extend of the half space geometry, $x>0$,
the Purcell factor rapidly drops as we increase the distance, for
a QE with a transition dipole moment along $z$.

In Fig.~\ref{fig:Fig04b} we observe that when the transition dipole
moment of the QE is oriented along $x$ the Purcell factor is less
enhanced right above the edge of the graphene half-space geometry
for the $n=0$ mode, which means that an $x$ oriented QE cannot excite
efficiently the edge SP mode right above the edge. As the QE is moved
towards the bulk of the graphene structure, $x<0$, we observe that
the Purcell factor for the edge mode is peaked at around $x=-5\,$nm.
At this position due to the $x$ transition dipole moment orientation
the QE can efficiently excite the edge SP mode. Increasing further
the distance from the edge of the half-space geometry the Purcell
factor drops for the $n=0$ edge mode due to the decoupling of the
near field of the QE to the edge SP mode. At these distances the $n=1$
bulk mode dominates, that is the main path of relaxation for the excited
QE. In striking contrast to the $z-$oriented transition dipole moment
of the QE, when the $x-$orientation is considered the Purcell factor
is enhanced when the QE is placed away of the half-space geometry,
$x>0$, and is peaked at a distance of $x=3\,$nm. Thus, when the
$x$ transition dipole moment of the QE is considered the edge SP
mode can be efficiently excited by placing a QE not exactly at the edge, but still close to the edge, of
the graphene half-space geometry.

\begin{figure*}
\subfloat[\label{fig:Fig05a}]{\includegraphics[width=0.45\textwidth]{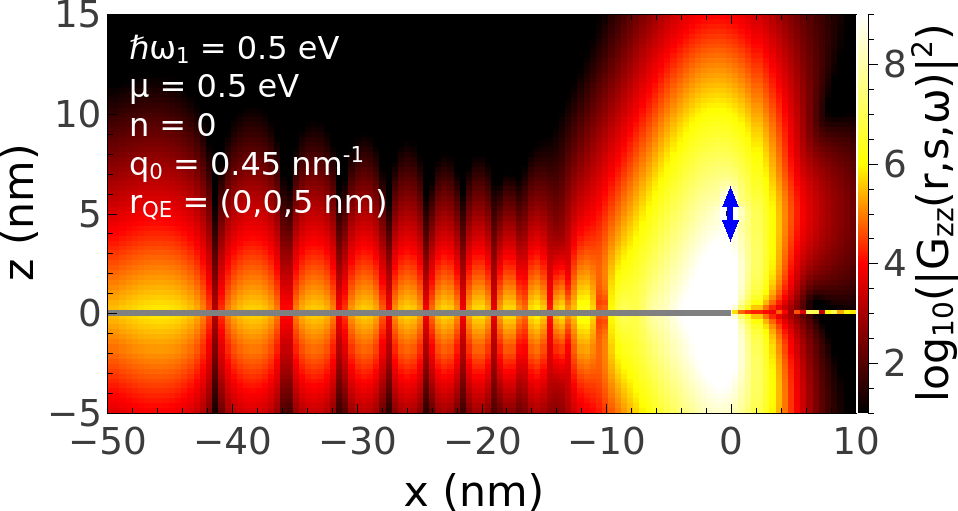}

}~~~~\subfloat[\label{fig:Fig05b}]{\includegraphics[width=0.45\textwidth]{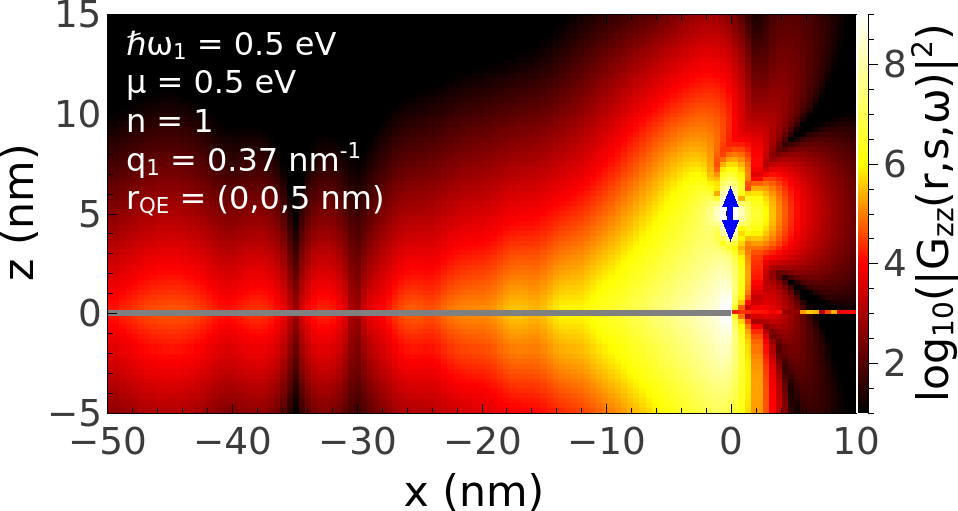}

}

\subfloat[\label{fig:Fig05c}]{\includegraphics[width=0.45\textwidth]{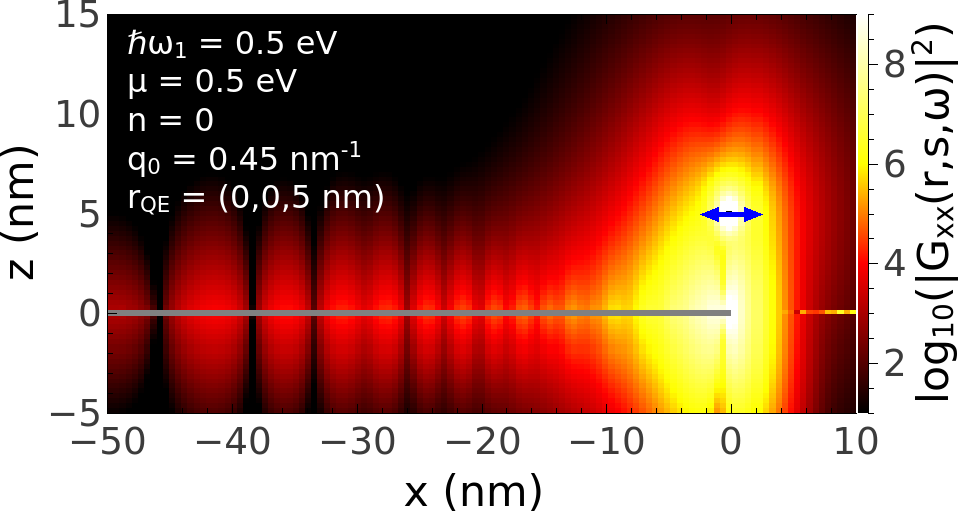}

}~~~~\subfloat[\label{fig:Fig05d}]{\includegraphics[width=0.45\textwidth]{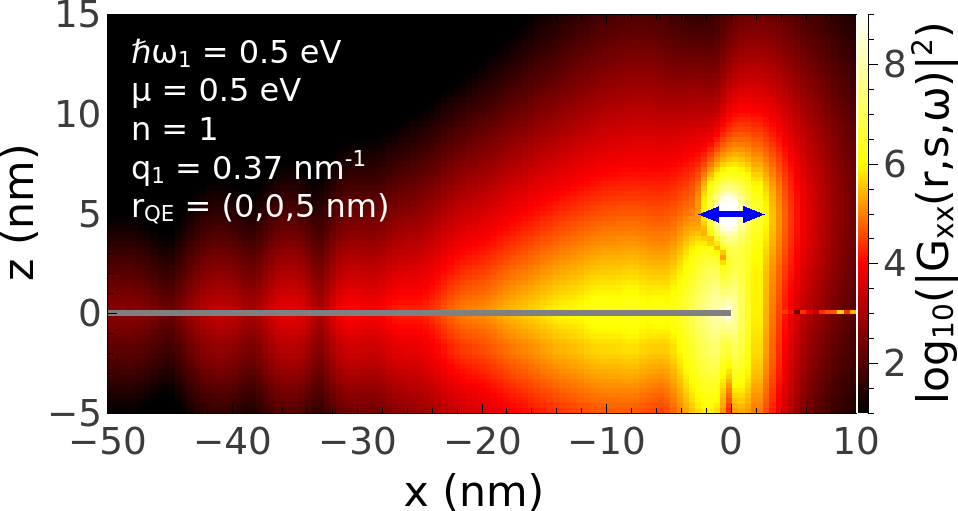}

}

\caption{Contour plots of the logarithm of the absolute value of the Green's
tensor for a QE placed at $\mathbf{r}=(0,0,5\,\text{nm})$ and emission
energy $\hbar\omega_{1}=0.5\,$ eV. The chemical potential of the graphene
half-space sheet is $\mu=0.5$ eV. Two orientations are considered (a,b)
$z$ and (c,d) $x$. Two modes are considered (a,c) $q_{0}=0.45\,\text{nm}^{-1}$
and (b,d) $q_{0}=0.37\,\text{nm}^{-1}$. \label{fig:Fig05}}

\end{figure*}

For better understanding the coupling of the QEs with the graphene
half space geometry, in Fig.~\ref{fig:Fig05} we present a contour
plot of the logarithm of the absolute value of the Green's tensor,
$\left|\mathfrak{G}\left(\mathbf{r,r}_{\text{QE}},\omega_{1}\right)\right|^{2}$,
created by a QE, which is placed at $\mathbf{r}_{\text{QE}}=(0,0,5\,\text{nm})$,
at the $x-z$ plane in the presence of the graphene half-space nanostructure.
Two different transition dipole moment orientations of the QE are
considered, $z$-oriented at Figs.~(\ref{fig:Fig05a}), (\ref{fig:Fig05b}) and
$x$-oriented at Figs.~(\ref{fig:Fig05c}), (\ref{fig:Fig05d}). The chemical
potential of graphene half-space is $\mu=0.5\,$eV and the emission
energy of the QE is $\hbar\omega_{1}=0.5\,$eV. Two resonance wave vectors
are considered $q_{0}=0.45\,\text{nm}^{-1}$ and $q_{1}=0.37\,\text{nm}^{-1}$
at Figs.~(\ref{fig:Fig05a}), (\ref{fig:Fig05c}) and (\ref{fig:Fig05b}), (\ref{fig:Fig05d}),
respectively. All color map scales in Fig.~\ref{fig:Fig05} are the
same, allowing the reader for a direct comparison.

In Fig. \ref{fig:Fig05a} we observe that a QE with a transition dipole
moment oriented along $z$ can launch an edge SP mode, $n=0$, on
the graphene half-space geometry efficiently. We observe that the
SP mode is confined to the graphene sheet tightly. The emission energy
of the QE is $\hbar\omega_{1}=0.5\,$eV ($\lambda=2480\,$nm) much larger
than the penetration depth of the SP field $\delta_{\text{SP}}^{n=0}=1/\text{Im}\left(k_{z}^{\text{SP}}\right)=2.2\,$nm,
as seen in Fig.~\ref{fig:Fig02c}, which matches the field extend
observed in Fig.~\ref{fig:Fig05a}. As  $\delta_{\text{SP}}^{n=0}/\lambda\sim 10^{-3}$, a three orders of magnitude
confinement of light is observed, in the perpendicular direction. Moreover, as the distance from the
edge towards the bulk graphene is increased the field intensity drops, and
the field is dissipated due to the materials losses, an effect that
is connected with the propagation length of the edge SP mode, where
$L_{\text{SP}}^{0}=585\,$nm. When the bulk resonance mode is
considered, for $n=1$ and $q_{1}=0.37\,\text{nm}^{-1}$, in Fig.~\ref{fig:Fig05b}
we observe that the field intensity is reduced, compared to the edge
mode $n=0$. Thus, the discontinuity of the graphene half-space geometry
can be used to launch an edge SP mode at the edge, that is extremely
confined on the sheet while it can travel hundreds of nanometers
with large field values, which potentially can be used to couple other
nearby QEs or nanostructures.

In Fig.~\ref{fig:Fig05c} we present a contour plot of the field
created by a $x$ oriented transition dipole moment QE, which is placed
exactly at the edge of the graphene half-space geometry, $\mathbf{r}_{\text{QE}}=(0,0,5\,\text{nm})$.
As someone would expect from the Purcell factor enhancement of Fig.~\ref{fig:Fig04b},
when the QE is placed at the edge of the half space is only poorly
coupled to the edge SP mode, $n=0$, thus the field distribution is
less profound than when a $z-$oriented transition dipole moment is
considered.
Nevertheless, we can see that the SP modes can be efficiently launched,
although their strength is smaller by two orders of magnitude compared
with the case a $z$ oriented QE is considered (Fig.~\ref{fig:Fig05a}).
When the in-plane wave vector $q$ matches the bulk mode, $q_{1}=0.37\,\text{nm}^{-1}$,
the field further drops due to the poor coupling with the bulk mode.
The main message obtained from Fig.~\ref{fig:Fig05} is that when
a QE is placed at specific areas of interest can excite SP modes, which can enhance the Purcell factor and create large field enhancements.
This control is active from the part of the QE, while for the case
where a metallic tip is used it cannot launch SP modes with an arbitrarily
small wave number due to the non-zero tip radius \cite{Fei2011a}.

\begin{figure*}
\subfloat[\label{fig:Fig06a}]{\includegraphics[width=0.45\textwidth]{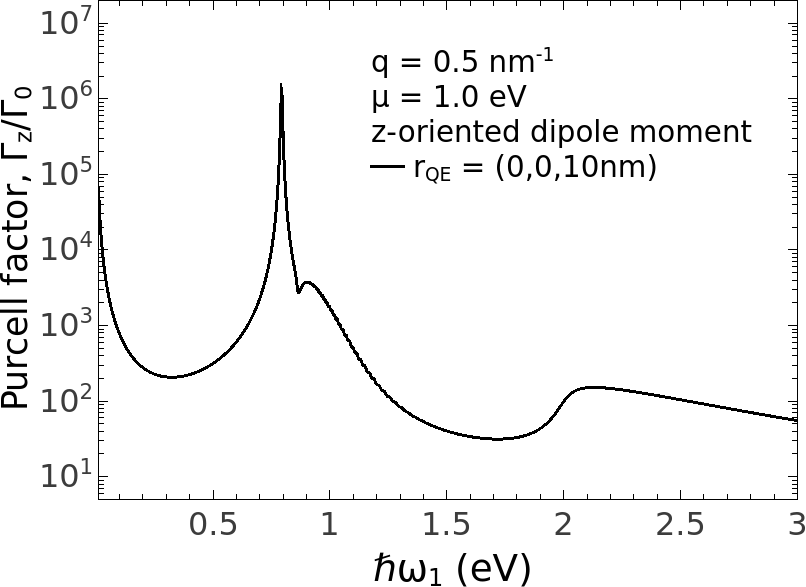}

}\subfloat[\label{fig:Fig06b}]{\includegraphics[width=0.525\textwidth]{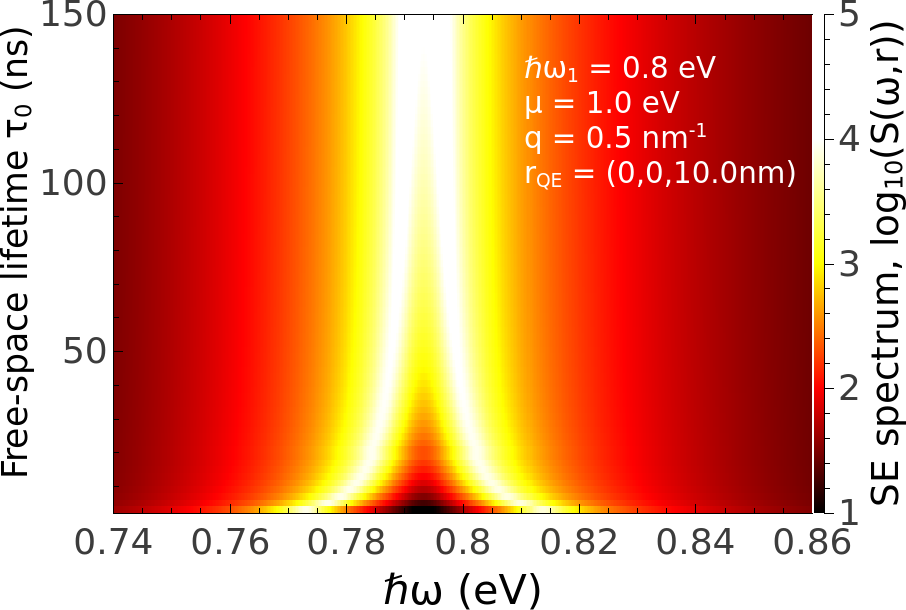}

}

\caption{A QE is placed at the edge of a graphene half-space geometry, with
a value of chemical potential $\mu=1\,$eV and in-plane wave vector
value $q=0.5\,\text{nm}^{-1}$, at $\mathbf{r}_{\text{QE}}=(0,0,10\,\text{nm})$. The transition dipole moment of the QE is $z$-oriented. (a)
Purcell factor of the QE.
(b) Contour plot of the logarithm of the SE spectrum, for $\hbar\omega_{1}=0.8\,$eV, and different
free-space lifetimes $\tau_{0}$. \label{fig:Fig06}}

\end{figure*}
From the analysis up to now it has become clear that the presented
enhancements of the Purcell factor of a QE are very large and thus
the combined interaction QE/half-space nanostructure may enter
in the strong coupling regime. In Fig.~\ref{fig:Fig06a} we present
the Purcell factor of the QE placed above the graphene half-space
geometry edge, at $\mathbf{r}=(0,0,10\,\text{nm})$ with a $z-$oriented
transition dipole moment. The graphene is kept at a chemical potential
value of $\mu=1\,$eV and the in-plane wave vector has a value
of $q=0.5\,\text{nm}^{-1}$. We observe that at $\hbar\omega_{1}=0.8\,$eV the Purcell factor becomes above $10^6$. As we have already analyzed
the Purcell factor is due to the excitation of the edge mode supported
by the graphene half-space nanostructure.

From Eq.~\ref{eq:04} it is clear that the value of the spectral density, for specific Purcell factor, depends on the free-space SE rate
$\tau_{0}=1/\Gamma_{0}(\omega_{1})$ of the QE. The chemical potential of the graphene is $\mu=1\,$eV
and the in-plane wave vector is $q=0.5\,\text{nm}^{-1}$. In Fig.~\ref{fig:Fig06b}
we present a contour plot of the logarithm of the SE spectrum of the QE near the graphene half-space nanostructure, Eq.~\ref{eq:05},
for different free-space lifetimes of the QE, $\tau_{0}$. The transition energy from the excited to the
ground state of the QE is at $\hbar\omega_{1}=0.8\,$eV ($\lambda=1550\,$nm)
which matches the highest value of the enhancement of the SE rate
of the QE, presented in Fig.~\ref{fig:Fig06b}. We observe that for
the free-space lifetime value $\tau_{0}=1\,$ns (typical for quantum dots) there is a Rabi-splitting
in the emission spectrum of the combined system which has a value
of $\hbar\Omega=80\,m$eV. As the value of the free-space lifetime,
$\tau_{0}$, is increased the splitting drops and at free-space lifetimes
above $120\,$ns disappears and there is a transition from the strong
to the weak coupling regime. Reducing the separation between the
QE and the graphene half space will increase the Purcell factor enhancement,
thus increasing the Rabi splitting. The same outcome we would have
if we had considered different QEs with smaller lifetimes $\tau_{0}$.

\section{Conclusions\label{sec:IV}}

In this article we investigated the SE properties of a QE near a graphene
half-space nanostructure in the weak and strong coupling regimes. A semi-analytical method was used to calculate the Green's
tensor at the electrostatic limit for the graphene half-space geometry,
where an expansion over Laguerre polynomials was applied. The graphene
half-space geometry supports edge and bulk SP modes, where by changing
the value of the chemical potential we can tune their properties, namely, the dispersion relation $\hbar\omega_{n}(q)$, the propagation
length $L_{n}$, and the penetration depth $\delta_{n}$ of the edge and
bulk SP modes, for $n=0$ and $1$, respectively.

We showed that when a QE is placed close to the edge of the half-space geometry
can efficiently excite the edge SP mode, and the Purcell factor
can take significant values. Different
transition dipole moments give different response to the enhancement
of the Purcell factors. When the QE has a $z-$transition dipole moment
orientation then the Purcell factor takes values larger than $10^8$, when the QE is placed
at $5\,$nm away from the graphene edge. Once the QE is placed
out of the surface of the graphene half-space, $x>0$, then a QE with
an $x$ transition dipole moment orientation dominates over a $z-$oriented
one, coupling more efficiently to the edge SP mode.

Furthermore, we showed that the QE when placed at the edge of the
graphene half-space geometry can launch edge and bulk SP modes without
the need to use tip-metallic excitation and having access to really
high wave vector values at high energies, relative to the chemical
potential value $\mu$. For all-optical
applications based on two dimensional materials this is quite important as someone can efficiently transmit or extract information from the device to the environment using QEs.

Finally, the Purcell factor enhancement of the QE can reach large
values and also has narrow spectral response, thus the interaction between the QE/graphene half-space nanostructure
enters the strong coupling regime. Rabi
splitting about $80\,m$eV is observed for a free space
lifetime of $\tau_{0}=1\,$ns, placed at a distance of $10\,$nm
above the graphene edge. The free-space emission energy is $\hbar\omega_{1}=0.8\,$eV
close to the telecommunication wavelength $\lambda=1550\,$nm, a wavelength range
that is very important for various applications.

As a next step in a similar line of research we will investigate the
strong coupling regime of the interaction between a pair of QE placed
in the vicinity of the graphene half-space geometry in the strong
coupling regime. This will be important for further understanding
how a practical quantum application can operate.

\appendix

\section{eigenfrequencies and eigenfunctions of the electrostatic potential\label{sec:Appendix-A}}

In this Appendix we present a method to calculate the Green's tensor
for the graphene half-space geometry in the electrostatic limit. The
electrostatic potential $\phi(\mathbf{r})$ is found by solving the
Poisson equation:
\begin{equation}
\nabla^{2}\phi(\mathbf{r})=-\frac{4\pi}{\varepsilon(\mathbf{r})}\rho(\mathbf{r}),\label{eq:A01}
\end{equation}
where $\varepsilon(\mathbf{r})$ is the dielectric permittivity and
$\rho(\mathbf{r})=\rho_{\parallel}(x,y)\Theta(-x)\delta(z)$ is the
charge distribution due to the electrons provided by the graphene
half space. Through out this paper we consider a constant value for
the dielectric permittivity $\varepsilon(\mathbf{r})=\varepsilon$.
This electrostatic problem has an obvious symmetry along the $y-$axis.
Thus, we can write the potential $\phi(\mathbf{r})$ and the charge
density $\rho(\mathbf{r})$ as: $\phi(\mathbf{r})=\phi(x)\phi_{\perp}(z)\text{exp}(iqy)$
and $\rho_{\parallel}(x,y)=\rho(x)\text{exp}(iqy)$. For $z\neq0$
and taking a Fourier transformation of Eq.~(\ref{eq:A01}) we extract
the following differential equation:
\begin{equation}
\left[\frac{d^{2}}{dz^{2}}-\left(k^{2}+q^{2}\right)\right]\phi_{\perp}(z)=0,\label{eq:A02}
\end{equation}
which has solutions of the form $\phi_{\perp}(q,z)=A_{\pm}\text{exp}\left(\mp\sqrt{k^{2}+q^{2}}\right)$.
Applying the relevant boundary conditions, that the potential $\phi$
is continuous and its normal derivative has a discontinuity of the
form:
\begin{equation}
\frac{\partial\phi(q,z)}{\partial z}\mid_{z=0^{+}}-\frac{\partial\phi(q,z)}{\partial z}\mid_{z=0^{-}}=-\rho(k),\label{eq:A03}
\end{equation}
the unknown coefficients have the form $A_{\pm}=\frac{1}{2\varepsilon\sqrt{k^{2}+q^{2}}}\rho(k)$.
Hence, the induced field is given by the expression:
\begin{equation}
\phi(\mathbf{r})=\frac{2e^{iqy}}{\varepsilon}\int_{-\infty}^{0}K(z,x,x^{\prime})\rho(x^{\prime})dx^{\prime},\label{eq:A04}
\end{equation}
where the kernel of the integral is given by the expression

\begin{equation}
K(z,x,x^{\prime})=\int_{-\infty}^{\infty}e^{i(x-x^{\prime})k}\frac{e^{-\sqrt{k^{2}+q^{2}}\left|z\right|}}{\sqrt{k^{2}+q^{2}}}.\label{eq:A05}
\end{equation}

We continue using Ohm's law, $\mathbf{J}=\sigma(\omega)\mathbf{E}$,
and the continuity equation, $i\omega\rho(x)=\mathbf{\nabla\cdot J}(x)$,
to find the expression
\begin{equation}
\phi_{\text{sur}}(x)=\frac{i\omega}{\sigma(\omega)q^{2}}\int_{-\infty}^{0}G(x,x^{\prime})\rho(x^{\prime})dx^{\prime},\label{eq:A06}
\end{equation}
where we have reintroduced $qx\to x$. $G(x,x^{\prime})$ is the Green's
function, which is solution of the equation
\begin{equation}
\left[\frac{d^{2}}{dx^{2}}-1\right]G(x,x^{\prime})=-\delta(x-x^{\prime}),\label{eq:A07}
\end{equation}
on the interval $(-\infty,0)$ and boundary condition
\begin{equation}
\left[\frac{d}{dx}G(x,x^{\prime})\right]_{x=0^{-}}=0.\label{eq:A08}
\end{equation}
The solution of Eq.~(\ref{eq:A07}) has the form
\begin{equation}
G(x,x^{\prime})=\frac{1}{2}\left(e^{x+x^{\prime}}+e^{-\left|x-x^{\prime}\right|}\right).\label{eq:A09}
\end{equation}
At $z=0$ we have to solve the following integral equation:\begin{widetext}
\begin{equation}
\omega^{2}\int_{-\infty}^{0}G(x,x^{\prime})\rho(x^{\prime})dx^{\prime}+\Omega^{2}\int_{-\infty}^{0}K(x,x^{\prime})\rho(x^{\prime})dx^{\prime}=-\frac{iq^{2}}{2}\omega\sigma(\omega)\phi_{ext}(x,x^{\prime}),\label{eq:A10}
\end{equation}
where $\Omega^{2}(\omega)=\frac{4i\omega\sigma(\omega)q}{\varepsilon}$
and $\phi_{ext}(x,x^{\prime})$ is the external excitation created
by a pseudo-potential. The only unknown quantity in the above equation
is the charge density $\rho(x)$, to proceed we will transform the
integral Eq.~(\ref{eq:A10}) to a matrix equation by expanding the
surface charge density, $\rho(x)$, using Laguerre polynomials\end{widetext}
\begin{equation}
\rho(x)=\sum_{j=0}^{\infty}c_{j}e^{x}L(-2x),\label{eq:A11}
\end{equation}
where $c_{j}$ is the set of coefficients that they need to be determined,
$L(-2x)$ are the Laguerre polynomials which satisfy the following
orthogonality condition
\begin{equation}
\int_{-\infty}^{0}e^{2x}L_{i}(-2x)L_{j}(-2x)=\frac{1}{2}\delta_{ij}.\label{eq:A12}
\end{equation}
Using the above expansion, the orthogonality condition, we get the
following matrix problem
\begin{equation}
\left(\omega^{2}\mathbf{G}-\Omega^{2}(\omega)\mathbf{K}\right)\mathbf{c}=-\frac{iq^{2}\omega\sigma(\omega)}{2}\mathbf{d},\label{eq:A13}
\end{equation}
where $\mathbf{c}$ is a column vector where its elements are the
expansion coefficients $c_{j}$, $\mathbf{d}$ are the expansion coefficients
of the dipole source with a $z$ dipole orientation and they are given
from the expression
\begin{equation}
d_{i}=2q^{2}\int_{-\infty}^{0}\frac{xL_{i}(-2x)e^{x}}{\left(x^{2}+z_{1}^{2}\right)^{3/2}}dx,\label{eq:A14}
\end{equation}
when a point dipole source oriented along $x$ is considered, and
\begin{equation}
d_{i}=2q^{2}\int_{-\infty}^{0}\frac{z_{1}L(-2x)e^{x}}{\left(x^{2}+z_{1}^{2}\right)^{3/2}}dx,\label{eq:A15}
\end{equation}
when the transition dipole moment is along $z$. The elements of the
matrices $\mathbf{G}$ and $\mathbf{K}$ are given by the expressions
\begin{equation}
G_{i,,j}=\frac{1}{8}\delta_{0,0}+\frac{1}{4}\delta_{i,j}-\frac{1}{8}\left(\delta_{i,i+1}+\delta_{i+1,i}\right)\label{eq:A16}
\end{equation}
and
\begin{equation}
K_{j,j+l}=\frac{-1}{\pi\left(2l+1\right)\left(2l-1\right)},\label{eq:A17}
\end{equation}
respectively \cite{Fetter1986}.

The induced part of the pseudopotential $\phi\left(\mathbf{r},\mathbf{r}^{\prime}\right)$
is given by the expression\begin{widetext}
\begin{equation}
\phi^{\text{ind}}\left(\mathbf{r},\mathbf{r}^{\prime}\right)=\frac{2e^{iy}}{\varepsilon(\mathbf{r})q}\sum_{j=0}^{\infty}(-1)^{j}c_{j}(z_{1})\int_{-\infty}^{\infty}dk\frac{e^{-\sqrt{k^{2}+1}z}}{\sqrt{k^{2}+1}}e^{ikx}\frac{\left(1+ik\right)^{j}}{\left(1-ik\right)^{j+1}},\label{eq:A18}
\end{equation}
where all the lengths are normalized to the in-plane wave vector $q$.
Also, we have used the integral relation
\begin{equation}
\int_{-\infty}^{0}dx\,e^{(1-ik)x}L_{j}(-2x)=(-1)^{j}\frac{\left(1+ik\right)^{j}}{\left(1-ik\right)^{j-1}}.\label{eq:A19}
\end{equation}

The induced electrostatic Green's tensor is given by the expression
\begin{equation}
\mathfrak{G}^{\text{ind}}(\mathbf{r,r^{\prime}},\omega)=\varepsilon_{0}c^{2}/(\omega p_{0})\mathbf{\nabla}\phi^{\text{ind}}(\mathbf{r,r^{\prime}}),\label{eq:A20}
\end{equation}
Then, the Green's tensor has the form
\begin{equation}
\mathfrak{G}_{zz}^{\text{ind}}\left(\mathbf{r},\mathbf{r}^{\prime},\omega\right)=\frac{2e^{iy}}{\varepsilon(\mathbf{r})}\sum_{j=0}^{\infty}\left(-1\right)^{j}c_{j}(\mathbf{r}^{\prime})\int_{-\pi}^{\pi}d\theta\,\frac{e^{i\tan\left(\theta\right)x}}{\cos\left(\theta\right)}e^{-\frac{z_{1}}{\cos\left(\theta\right)}}e^{i(2j+1)\theta},\label{eq:A21}
\end{equation}
which is connected with the $z$ component of the electric field created
by a dipole source oriented along $z$, and
\begin{equation}
\mathfrak{G}_{xx}^{\text{ind}}\left(\mathbf{r},\mathbf{r}^{\prime},\omega\right)=\frac{-2ie^{iy}}{\varepsilon(\mathbf{r})}\sum_{j=0}^{\infty}\left(-1\right)^{j}c_{j}(\mathbf{r}^{\prime})\int_{-\pi}^{\pi}d\theta\,e^{i\tan\left(\theta\right)x}\tan\left(\theta\right)e^{-\frac{z_{1}}{\cos\left(\theta\right)}}e^{i(2j+1)\theta},\label{eq:A22}
\end{equation}
\end{widetext}which is connected with the $x$ component of the electric
field created by a dipole source oriented along $x$.

\section{Graphene conductivity\label{sec:Appendix-B}}

The optical response of graphene is given by the value of the in-plane
conductivity, $\sigma$, in the random phase approximation \cite{Jablan2009,Falkovsky2008}.
This quantity is mainly determined by electron-hole pair excitations,
which can be divided into intraband and interband transitions. It
depends on the chemical potential $\mu$, the temperature $T$, and
the scattering energy $E_{S}$ values as
\begin{equation}
\sigma=\sigma_{\text{intra}}+\sigma_{\text{inter}},\label{eq:A23}
\end{equation}
where the intraband and interband contributions are given by \cite{Wunsch2006},
\begin{widetext}

\begin{equation}
\sigma_{\text{intra}}=\frac{2ie^{2}k_{B}T}{\hbar\pi(\hbar\omega+i\hbar/\tau)}\ln\bigg[2\cosh\bigg(\frac{\mu}{2k_{B}T}\bigg)\bigg],\label{eq:A24a}
\end{equation}
\begin{equation}
\sigma_{\text{inter}}=\frac{e^{2}}{4\hbar}\bigg[\frac{1}{2}+\frac{1}{\pi}\arctan\bigg(\frac{\hbar\omega-2\mu}{2k_{B}T}\bigg)-\frac{i}{2\pi}\ln\frac{(\hbar\omega+2\mu)^{2}}{(\hbar\omega-2\mu)^{2}+(2k_{B}T)^{2}}\bigg].\label{eq:A24b}
\end{equation}
\end{widetext}The intraband term $\sigma_{\text{intra}}$ describes
a Drude model response, corrected for scattering by impurities through
a term containing $\tau$, the relaxation time. Throughout this paper
we consider room temperature $T=300\,\text{K}$ and a value of the
relaxation time of $\tau=1\,\text{ps}$ and we vary the value of chemical
potential, $\mu$ \cite{Novoselov2004}.

\bibliographystyle{prsty}

\end{document}